\documentclass[conference]{IEEEtran}

\usepackage{times}
\usepackage{cite}
\usepackage[numbers]{natbib}
\usepackage{multicol}
\usepackage[bookmarks=true]{hyperref}
\usepackage{amsmath,amssymb,amsfonts,amsthm}
\usepackage[linesnumbered,ruled,vlined]{algorithm2e}
\usepackage{algpseudocode}
\usepackage{graphicx}
\usepackage{textcomp}
\usepackage{xcolor}
\usepackage{subcaption}
\usepackage{stfloats}
\usepackage{bbm}
\usepackage{yhmath}

\usepackage{mymacros}

\SetKwInput{KwInput}{Input}                
\SetKwInput{KwOutput}{Output}              
\SetKw{Break}{break}

\begin{document}

\title{Adversarial optimization leads to over-optimistic security-constrained dispatch, but sampling can help
}
\pdfinfo{
   /Author (Charles Dawson, Chuchu Fan)
   /Title  (Accelerating failure prediction and mitigation using inference and automatic differentiation)
   /CreationDate (D:20220101120000)
   /Subject (verification, design optimization, differentiable simulation)
   /Keywords (verification, design optimization, differentiable simulation)
}

\author{\IEEEauthorblockN{Charles Dawson}
\IEEEauthorblockA{\textit{Dept. of Aeronautics and Astronautics} \\
\textit{Massachusetts Institute of Technology}\\
Cambridge, USA \\
\texttt{cbd@mit.edu}}
\and
\IEEEauthorblockN{Chuchu Fan}
\IEEEauthorblockA{\textit{Dept. of Aeronautics and Astronautics} \\
\textit{Massachusetts Institute of Technology}\\
Cambridge, USA \\
\texttt{chuchu@mit.edu}}
}

\maketitle

\begin{abstract}
To ensure safe, reliable operation of the electrical grid, we must be able to predict and mitigate likely failures. This need motivates the classic security-constrained AC optimal power flow (SCOPF) problem. SCOPF is commonly solved using adversarial optimization, where the dispatcher and an adversary take turns optimizing a robust dispatch and adversarial attack, respectively. We show that adversarial optimization is liable to severely overestimate the robustness of the optimized dispatch (when the adversary encounters a local minimum), leading the operator to falsely believe that their dispatch is secure.

To prevent this overconfidence, we develop a novel adversarial sampling approach that prioritizes diversity in the predicted attacks. We find that our method not only substantially improves the robustness of the optimized dispatch but also avoids overconfidence, accurately characterizing the likelihood of voltage collapse under a given threat model. We demonstrate a proof-of-concept on small-scale transmission systems with 14 and 57 nodes.
\end{abstract}

\begin{IEEEkeywords}
SCOPF, adversarial optimization, MCMC
\end{IEEEkeywords}
\IEEEpeerreviewmaketitle

\section{Introduction}

\begin{figure*}[t]
    \centering
    \includegraphics[width=\linewidth]{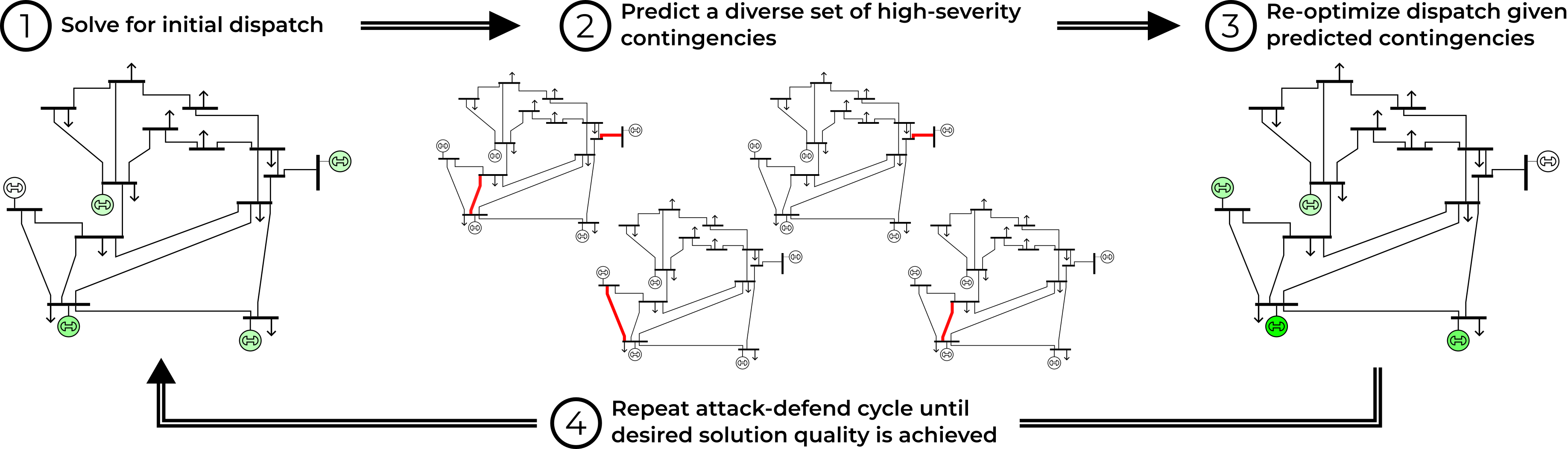}  
    \caption{An overview of our approach to contingency prediction and mitigation for SCOPF. (1) We solve for an initial ACOPF dispatch using stochastic optimization. (2) We attack the proposed dispatch, using a sampling-based algorithm to find a diverse set of contingencies with high expected severity. (3) We attempt to re-solve the ACOPF problem to mitigate the predicted contingencies, yielding a revised dispatch. (4) We repeat the attack-defend cycle until the desired solution quality is achieved. Red lines show high expected-severity outages for the initial dispatch. Green indicate utilization fraction of each generator in the initial and improved dispatch.}
    \label{fig:headline}
\end{figure*}


If we are to ensure a reliable supply of electricity, we must be able to predict and (more importantly) mitigate failure modes in complex electrical energy systems.
The ability to predict high-risk failures or contingencies is important for a number of reasons. When planning a new network, the system operator might wish to evaluate proposed grid designs by the kinds of failures likely to occur in each. Once a network has been constructed, an operator might use knowledge of likely failure modes to prioritize spending on infrastructure hardening and cybersecurity measures. During routine operation, this knowledge helps operators route electricity through the network in a way that is robust to likely failures (solving the security-constrained AC optimal power flow problem, or SCOPF). Finally, leading up to extreme events like hurricanes, an operator might use this knowledge to prepare, stationing repair crews near the site of likely failures.

This SCOPF problem is commonly solved using \textit{adversarial optimization}, where the dispatcher and adversary take turns optimizing a dispatch and attack, respectively. Unfortunately, this sort of optimization, which is inherently local, is liable to overestimate the robustness of the resulting dispatches. To address this problem, we develop a sampling-based tool for automatically predicting and mitigating contingencies in a transmission and distribution network. Specifically:
\begin{itemize}
    \item We propose a novel sampling-based algorithm using gradient-based Markov chain Monte Carlo (MCMC) to efficiently find high-cost contingencies,
    \item We combine contingency prediction with a contingency-mitigation algorithm in an adversarial MCMC framework to solve SCOPF, and
    \item We demonstrate that our combined prediction-and-mitigation approach (shown in Fig.~\ref{fig:headline}) yields robust solutions to SCOPF without the overconfidence typical of traditional optimization-based approaches.
\end{itemize}

\section{Related Work}\label{related_work}

The problem of finding and mitigating contingencies in a power transmission system has been considered in various forms, most notably the security-constrained AC optimal power flow (SCOPF) problem, for at least forty years~\cite{cainHistoryOptimalPower2012}. However, the computational difficulty of this problem means that it has not yet been fully solved (as evidenced by the recent ARPA-e Grid Optimization challenge seeking the development of new SCOPF algorithms~\cite{GridOptimizationCompetition}). 

A large number of approaches have been developed to predict and mitigate single-point contingencies (so-called $N-1$ failures), but these techniques cannot account for risks due to multiple faults~\cite{yanConvexThreeStageSCOPF2021,bazrafshanComputationallyEfficientSolutions2020,demagalhaescarvalhoSecurityConstrainedOptimalPower2018}. Some techniques exist to handle the more general $N-k$ case; most of these approaches use an adversarial optimization scheme where an upper layer solves for the optimal dispatch while lower layers search for a worst-case contingency~\cite{hongNKConstrainedComposite2017,wangTwostageRobustOptimization2013,dontiAdversariallyRobustLearning2021}.

There are two primary issues with adversarial optimization as proposed in the existing literature. First, adversarial optimization methods with nonlinear power flow constraints are \textit{inherently local}; while they may converge to a locally optimal contingency, there is no guarantee that they will find the true worst case. Second, these approaches tend to treat the number of failures $k$ as a hyperparameter specified by the user~\cite{dontiAdversariallyRobustLearning2021}; instead, we would prefer a method that dynamically adjusts $k$ to find contingencies that are both likely and high-severity.

In this paper, we tackle both of these problems by reformulating the SCOPF problem as a probabilistic inference problem (taking inspiration from work on rare-event prediction~\cite{okellyScalableEndtoEndAutonomous2018,rubinoIntroductionRareEvent2009} and cyberphysical system verification~\cite{corsoSurveyAlgorithmsBlackBox2021}). The benefits of this approach are twofold. First, our formulation admits powerful stochastic algorithms that can avoid local minima. Second, taking a probabilistic perspective allows us to consider the \textit{likelihood} of failures in addition to their severity, searching for failures that maximize the expected severity of failures. This allows us to weight low-probability events with high severity similarly to more-common events with only moderate severity.

It is important to note that our approach is similar to that in~\cite{demagalhaescarvalhoSecurityConstrainedOptimalPower2018}, which formulates the $N-1$ SCOPF problem as a rare-event prediction problem, but \cite{demagalhaescarvalhoSecurityConstrainedOptimalPower2018} assumes that a single fixed contingency is specified by the user, and thus it does not solve the full prediction-and-mitigation problem we consider here.

\section{SCOPF Problem Statement}\label{prelim}

Simply put, SCOPF aims to solve for a dispatch (power injections and voltages) that ensures that bus voltages are stable even in the event of transmission outages.
Formally, we define a \textit{dispatch} $x = (P_g, |V|_g, P_l, Q_l)$ to include real power generation $P_g$ and AC voltage amplitude $|V|_g$ at each generator and real and reactive power demand $P_l$, $Q_l$ at each load (some loads, like vehicle charging, are capable of varying in response to network conditions; for non-dispatchable loads these demands will be fixed). To model transmission outages, we introduce a scalar strength $y_i \in \R$ of each transmission line in the network; the admittance of each line is given by $\sigma(y_i) Y_{i, nom}$ where $\sigma$ is the sigmoid function and $Y_{i, nom}$ is the nominal admittance of the line. When $y_i \gg 0$, the line operates at its nominal admittance, but when $y_i \leq 0$ the line is impaired. We refer to the vector of line strengths as $y = [y_1, \ldots, y_n]$, where each $y$ represents a contingency scenario.

Given a dispatch $x$ and contingency $y$, we solve the nonlinear AC power flow equations~\cite{dontiAdversariallyRobustLearning2021, dontiDC3LearningMethod2021} for the AC voltage amplitudes $|V|$ at each non-generator bus, the voltage phase angles $\theta$ at each non-slack bus, and the net real and reactive power injections $(P, Q)$ at each bus. We follow the 2-step method described in~\cite{dontiDC3LearningMethod2021} where we first solve for the voltage magnitudes and angles using a nonlinear equation solver and then compute the reactive power injection from each generator and the power injection from the slack bus. We then assign a scalar \textit{severity} to each dispatch-contingency pair that combines the economic cost of generation $c_g$ (a quadratic function of $P_g, P_l, Q_l$) with the total violation of constraints on generator capacities, load requirements, and voltage amplitudes:
\begin{align}
    S(x, y) =& \sum_{g \in \mathcal{G}} c_g + v(P_g, \underline{P}_{g}, \bar{P}_{g}) + v(Q_g, \underline{Q}_{g}, \bar{Q}_{g}) \\
    & + \sum_{l \in \mathcal{L}} v(P_l, \underline{P}_{l}, \bar{P}_{l}) + v(Q_l, \underline{Q}_{l}, \bar{Q}_{l}) \\
    & + \sum_{i \in \mathcal{B}} v(|V_i|, |\underline{V}_i|, |\bar{V}_i|) \label{eq:scopf_cost}
\end{align}
where $\underline{\cdot}$ and $\bar{\cdot}$ denote min/max limits,
$v(x, \underline{x}, \bar{x}) = L\pn{[x - \bar{x}]_+ + [\underline{x} - x]_+}$ denotes the violation of a given constraint,
$L$ is a large penalty coefficient, $[\cdot]_+ = \max(\cdot, 0)$ is a hinge loss, and $\mathcal{G}$, $\mathcal{L}$, and $\mathcal{B}$ are the sets of generators, loads, and all buses, respectively.

Severity captures both the reduced economic efficiency resulting from an outage along with the (much more heavily-weighted) violation of network constraints. However, simply searching for contingencies that maximize severity would bias our search towards highly-unlikely multiple-fault scenarios (e.g. taking out all transmission lines in a network would lead to a high-severity failure, but this scenario is unlikely). To remedy this, we introduce the notion of \textit{risk-adjusted severity}:
\begin{align}
    S_r(x, y) = S(x, y) + \log(p_0(y))
\end{align}
where $p_0(y)$ is the prior probability of the contingency $y$ (e.g. specifying that each line has an independent 5\% probability of failure). This prior probability density may be derived from manufacturer data or estimated from historical performance, and it may capture correlations between transmission lines (e.g. if neighboring lines are likely to fail at the same time due to shared weather events).

Given this context, we frame SCOPF as a two-player zero-sum game over risk-adjusted severity played out between the dispatcher and an attacker:
\begin{align}
    x^* = \argmin_{x \in \cX} \max_{y \in \cY} S_r(x, \phi) \label{minmax}
\end{align}
Min/max problems like~\eqref{minmax} are often solved using alternating gradient-descent, as in~\cite{dontiAdversariallyRobustLearning2021,dawsonRobustCounterexampleguidedOptimization2022}, but in the next section we will show how this approach can lead to false negatives: cases when the optimization returns a dispatch that it believes to be safe while failing to discover likely high-severity failures.

\section{Issues with adversarial optimization}

A common strategy for min/max problems like~\eqref{minmax} is \textit{adversarial optimization}, where we alternate between optimizing the dispatch $x$ (using either gradient descent or a domain-specific solver) and optimizing the contingency $y$ (often using gradient descent, as in~\cite{dontiAdversariallyRobustLearning2021}. Iterating until both the dispatch and the contingency converge leads to a local Nash equilibrium between the dispatcher and the adversary~\cite{dawsonRobustCounterexampleguidedOptimization2022}, meaning that neither player can achieve a better result by making small modifications to their solution. However, this alternating optimization process is inherently local; it cannot guarantee that the resulting solution is globally optimal or that it has identified the true worst-case contingency.

Even when we extend adversarial optimization to use a population of multiple likely contingencies $\set{y^1, \ldots, y^{n_y}}$, optimizing each contingency independently from different initial conditions, there are no guarantees that we have accurately covered the space of contingencies. Worse, there is a risk that the dispatcher might \textit{overfit} to the predicted contingencies, resulting in a dispatch that is robust to the predicted set of contingencies but performs poorly otherwise. This results in false optimism: the dispatcher might think they have successfully mitigated all likely contingencies, but they are in fact stuck in a local optimum and have failed to discover high-severity failures that are ``just over the horizon.''

\begin{figure}[tb]
    \centering
    \includegraphics[width=\linewidth]{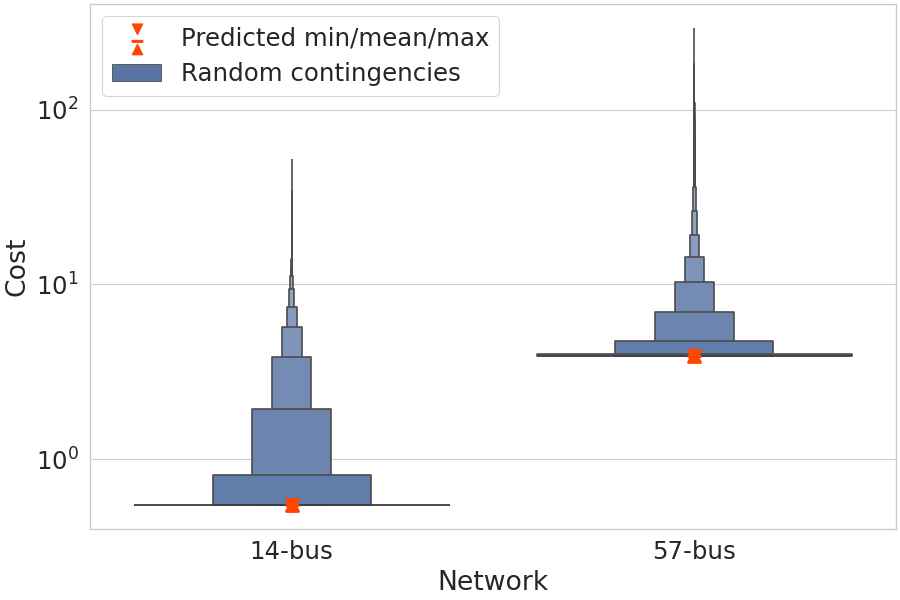}
    \caption{Adversarial optimization achieves good best- and average-case performance, but it predicts a set of contingencies that do not accurately reflect the likely worst-case. Although none of the contingencies predicted by the adversary lead to failure, 17.7\% of randomly sampled contingencies result in violation of voltage or power limits.}
    \label{fig:adv_opt_failure}
\end{figure}

We can see this behavior in practice on the 14- and 57-bus IEEE test networks. If we optimize a set of 10 likely contingencies and alternate between updating the dispatch to reduce average severity and updating the contingencies to maximize risk-adjusted severity, the process converges to a solution where the optimized dispatch is robust to all 10 predicted contingencies (achieving the same low cost on all of these contingencies). This might lead a system operator to believe that their solution is robust, but when we stress-test this dispatch with $10^6$ randomly-sampled contingencies, as shown in Fig.~\ref{fig:adv_opt_failure}, we see that the average system performance is much worse than the predicted contingencies might lead us to believe. In fact, the 14-bus dispatch encounters failures that are worse than the predicted contingencies in 17.7\% of cases.

\section{Avoiding false optimism using MCMC}\label{approach}

In this section, we introduce our approach to avoiding these false negatives by taking a sampling-based (rather than optimization-based) approach to solving the SCOPF problem. We will first describe how a generic optimization problem can be reduced to a sampling problem that can be solved using Markov chain Monte Carlo (MCMC) algorithms. We will then use this reduction to both predict contingencies, by solving the inner maximization of~\eqref{minmax}, and then to SCOPF as a whole.

\subsection{Sampling using Markov chain Monte Carlo}\label{sampling_as_optimization}

\begin{figure}[tb]
    \centering
    \includegraphics[width=\linewidth]{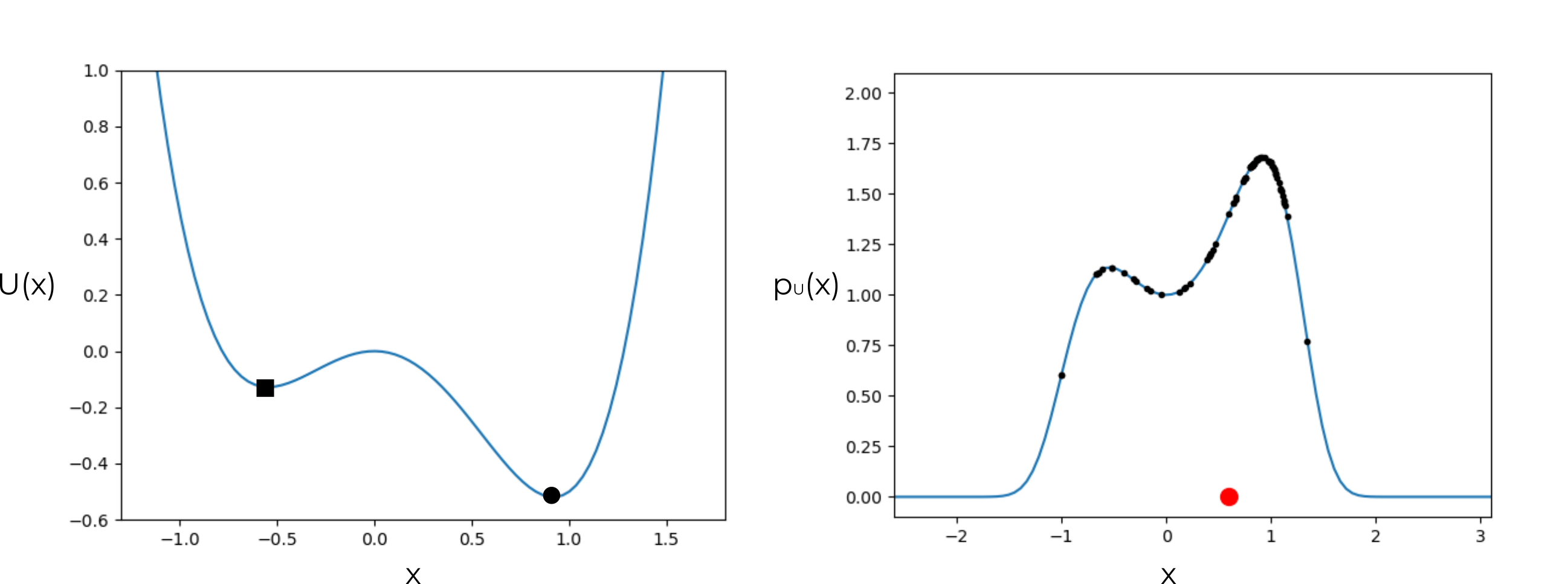}
    \caption{(Left) The cost landscape for an example minimization problem with 2 local minima (the square and circle). If improperly initialized, a local gradient-based optimizer will fail to converge to the global minimum. (Right) Converting this cost to a probability density and sampling from the resulting distribution allows us to escape the local minimum even when initialized far to the left (the red dot shows the mean of 100 samples).}
    \label{fig:sampling_as_optimization}
\end{figure}

Consider the generic unconstrained optimization problem:
\begin{align}
    x^* = \argmin U(x)
\end{align}
The cost landscape for an example $U(x) = x^4 - 0.5 x^3 - x^2$ is shown on the left in Fig.~\ref{fig:sampling_as_optimization}. This cost function has local minima at $x = -0.544$ and $x = 0.919$ (the latter represents the true global minimum). If initialized with $x < 0$, a local gradient-based optimizer will naturally converge to the sub-optimal local minimum at $x = -0.544$ (the square in Fig.~\ref{fig:sampling_as_optimization}), missing the global minimum (marked by a circle).

This behavior is a hallmark of gradient-based optimization; however, if we re-frame this optimization as a the problem of sampling from an appropriately defined probability distribution, we can exploit efficient gradient-based algorithms that are not prone to getting stuck in local optima~\cite{maSamplingCanBe2019}. In particular, let us define the probability density function (often referred to simply as a likelihood):
\begin{align}
    p(x) \propto e^{-U(x)}
\end{align}
Note that we are not required to normalize this distribution (which would be intractable in general). This likelihood is shown on the right in Fig.~\ref{fig:sampling_as_optimization}; taking the average of 100 samples drawn from this density yields an approximation of the global optimum (the red dot in Fig.~\ref{fig:sampling_as_optimization}; a much closer approximation could be found by simply taking the sample with the highest likelihood).

Of course, sampling from $p(x)$ for an arbitrarily complicated $U$ is non-trivial, but a class of algorithms known as Markov chain Monte Carlo (MCMC) exist to do precisely this~\cite{geyerIntroductionMarkovChain2011}. The key insight behind MCMC is to define a Markov process that uses the value of $p$ (and optionally its gradient) to propose a sequence of states that approximate the distribution $p(x)$. Thus, after a sufficient "warm-up" period, this Markov chain will randomly explore the state space in a way that makes its probability of visiting any particular state proportional to that state's likelihood $p(x)$.

There is a rich history of research into MCMC methods (see~\cite{geyerIntroductionMarkovChain2011} for an introduction), including both practical methods for accelerating sampling and theoretical analysis of convergence rates~\cite{maSamplingCanBe2019}. Of particular note is the Metropolis-adjusted Langevin algorithm (MALA), which combines gradient information with a biased random walk to quickly converge to the target distribution~\cite{maSamplingCanBe2019}. A summary of MALA is given in Algorithm~\ref{alg:mala}; the key idea is to balance the effects of diffusion (a Gaussian random walk through the state space) with a gradient-driven drift towards higher-likelihood regions and an accept-reject step to incorporate information about the relative value of $p$ in different regions. Note that Algorithm~\ref{alg:mala} only requires evaluating the gradient of the logarithm of $p$ and the ratio of $p$ at two different points, so $p$ need only be known up to a constant normalizing factor.

\begin{algorithm}
\caption{Metropolis-adjusted Langevin algorithm}\label{alg:mala}
\DontPrintSemicolon
    \KwInput{Initial $x_0$, steps $K$, stepsize $\tau$, density $p(x)$.}
    \KwOutput{A sample drawn from $p(x)$.}
    \For{$i = 1, \ldots, K$}
    {
        Sample $\eta \sim \cN(0, 2\tau I)$ \Comment{Gaussian noise}\;
        $x_{i+1} \gets x_i + \tau \nabla \log p(x_i) + \eta$ \Comment{Propose next state}\;
        $P_{accept} \gets \frac{p(x_{i+1}) e^{-||x_i - x_{i+1} - \tau \nabla \log p(x_{i+1})||^2 / (4\tau)}}{p(x_{i}) e^{-||x_{i+1} - x_{i} - \tau \nabla \log p(x_{i})||^2 / (4\tau)}}$ \;
        With probability $1 - \min(1, P_{accept})$:\;
        \hspace{2em}$x_{i+1} \gets x_{i}$ \Comment{Accept/reject proposal}\;
    }
    \KwRet{$x_K$}
\end{algorithm}

\subsection{Contingency prediction}\label{prediction}

We will now demonstrate how this optimization-to-sampling reduction can be applied to automatically predict contingencies with high risk-adjusted severity while avoiding the false negatives encountered with adversarial optimization. Recall that the problem of predicting contingencies for a given dispatch $x$ can be formulated as the optimization
\begin{align}
    y^* = \argmax_{y \in \cY} S_r(x, y)
\end{align}
This problem can be reduced to a sampling problem with the severity-adjusted distribution
\begin{align}
    y \sim p(y | x) \propto e^{S_r(x, y)} = p_0(y) e^{S(x, y)}
\end{align}

In plain language, sampling from this distribution will yield contingencies that have both high severity and high prior likelihood. By balancing the prior probability against the severity, we avoid having to specify a fixed number of failures; highly unlikely multi-point failures will naturally be de-prioritized unless the severity of the outage compensates for the low prior likelihood.

\subsection{Contingency mitigation}\label{mitigation}

Once we have the ability to sample contingencies with high risk-adjusted severity, the next step is to use that knowledge to preemptively modify the dispatch to reduce the impact of the predicted contingencies. Recall that the search for a robust dispatch to mitigate possible contingencies can be framed as the min/max optimization problem~\eqref{minmax}. As in Sections~\ref{sampling_as_optimization} and~\ref{prediction}, our goal is to reduce this min/max problem to a sampling problem. To do this, we take inspiration from multi-level optimization approaches to solving min/max problems~\cite{dontiAdversariallyRobustLearning2021,dawsonRobustCounterexampleguidedOptimization2022} to develop a sequential Monte Carlo (SMC) framework.

Our approach is shown in Algorithm~\ref{alg:smc}. Walking through this algorithm, it maintains a population of $n_x$ possible dispatches and $n_y$ possible contingencies, evolving each population in response to the other. The dispatch population is initialized uniformly at random, while the contingency population is initialized i.i.d. from the prior $p_0(y)$. The algorithm then proceeds for $N$ rounds. In each round, we first update our dispatch population to perform better against the current set of predicted contingencies, then update the contingencies to perform better against the current best dispatch. In the first round, contingencies are predicted solely based on the prior, but in subsequent rounds the contingency population will evolve to contain scenarios with higher risk-adjusted severity. Both the contingencies and the dispatches are updated by running MALA (Algorithm~\ref{alg:mala}) on an appropriate likelihood. The likelihood used to update the contingencies is the same used for contingency prediction in Section~\ref{prediction}, where the goal is to maximize the risk-adjusted severity of the contingency for the current best dispatch. The likelihood used to update the dispatch is designed to minimize the expected severity averaged over the current population of contingencies

\begin{algorithm}
\caption{SMC approach to contingency prediction and mitigation}\label{alg:smc}
\DontPrintSemicolon
    \KwInput{Dispatch and contingency population sizes $n_x$, $n_y$, alternating steps $N$, substeps $K$, stepsize $\tau$.}
    \KwOutput{A robust dispatch $x^*$ and a set of likely worst-case contingencies $\set{y_1, \ldots, y_{n_y}}_K$.}
    Initialize dispatches $[x]_0 = \set{x_1, \ldots, x_{n_x}}_0$ sampled uniformly from $\cX$\;
    Initialize contingencies $[y]_0 = \set{y_1, \ldots, y_{n_y}}_0$ sampled from the prior $p_0(y)$\;
    \For{$i = 1, \ldots, N$}
    {
        \Comment{Update dispatches using predicted contingencies}\;
        Define $U_{x, i}(x) = \sum_{y \in [y]_{i-1}} S(x, y) / n_y$\;
        $[x]_i \gets \rm{MALA}([x]_{i-1}, K, \tau, p(x) \propto e^{-U_{x, i}(x)})$\;
        \Comment{Update contingencies against new best dispatch}\;
        Define $U_{y, i}(y) = -\min_{x \in [x]_{i}} S_r(x, y)$ \;
        $[y]_i \gets \rm{MALA}([y]_{i-1}, K, \tau, p(x) \propto e^{-U_{y, i}(x)})$\;
    }
    $x^* \gets \argmin_{x \in [x]_{N}} U_{x, K}(x)$ \Comment{Choose best dispatch}
    \KwRet{$x^*$, $[y]_n = \set{y_1, \ldots, y_{n_y}}_N$}
\end{algorithm}

\begin{figure}[tb]
    \centering
    \includegraphics[width=\linewidth]{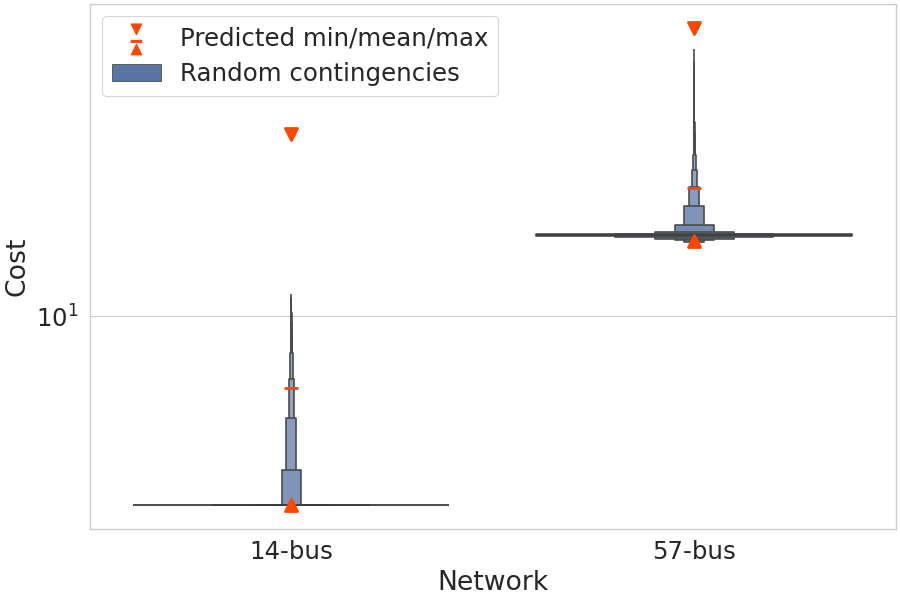}
    \caption{In contrast to adversarial optimization (Fig.~\ref{fig:adv_opt_failure}), our method finds a dispatch that is highly robust, with much lower variance in cost across randomly-sampled contingencies. Moreover, our method predicts likely contingencies that cover the space of possible failures (none of the randomly sampled contingencies were worse than the predicted contingencies).}
    \label{fig:smc_success}
\end{figure}

Compared to a strict optimization-based approach, Alg.~\ref{alg:smc} leads to much better exploration of the failure space and generates a more diverse set of contingencies. Fig.~\ref{fig:smc_success} shows the robustness of the dispatch found using our method on the 14- and 57-bus examples (running in \SI{142}{s} and \SI{1438}{s}, respectively). The dispatch found using our method is more conservative than that found using adversarial optimization (achieving a higher average economic cost), but it is substantially more robust (with much lower variance observed on the test set). Moreover, while adversarial optimization yields a set of contingencies that do not accurately predict performance on the test set, our method predicts a diverse set of contingencies that accurately cover the true performance of the system. These results show that our sampling-based method avoids the false optimism of gradient-based optimization, and we do not observe any failures in the test set that exceed the predicted worst-case.

\section{Comparison of failure modes}

We have shown that standard adversarial optimization tends to overestimate the robustness of its solution, while our method yields more robust dispatches and avoids this overconfidence. In addition to analyzing the distribution of costs observed for each method (Figs.~\ref{fig:adv_opt_failure} and~\ref{fig:smc_success}), we can also compare the types of failures seen with each method.

\begin{figure}[tb]
    \centering
    \includegraphics[width=\linewidth]{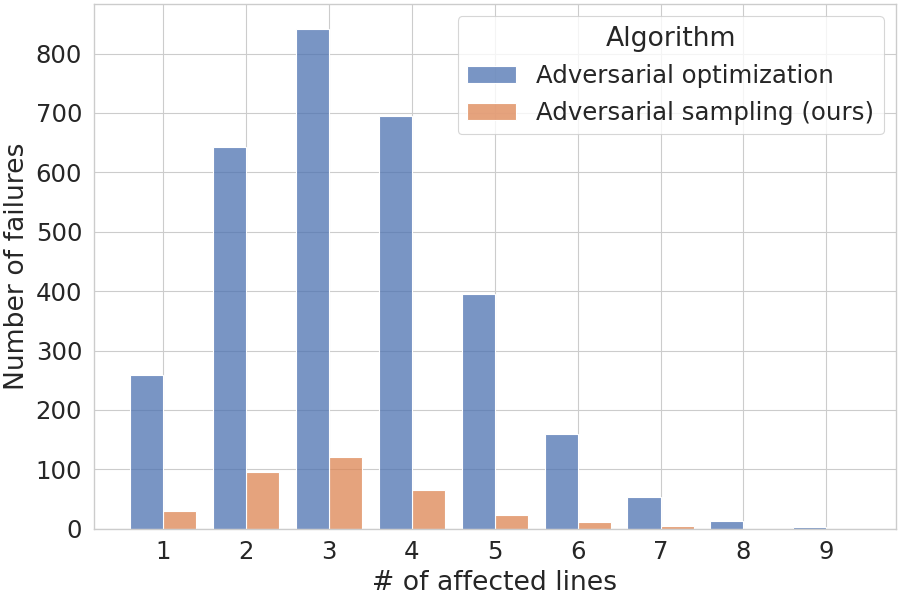}
    \caption{The distribution of outages that cause violation of voltage/power limits for the dispatch found using each method.}
    \label{fig:failure_comparisons}
\end{figure}

Fig.~\ref{fig:failure_comparisons} shows the distribution of line outages that induce a violation of power or voltage limits at any bus in the 14-bus test network (in terms of the number of lines operating at less than 90\% rated capacity). Across a test set of one million randomly-sampled network conditions, we observe a stark difference between standard adversarial optimization and our method. We see that at any given outage level, our method experiences a substantially reduced number of failures (9x fewer single outages, 7x fewer double outages, 7x fewer triple outages, 11x fewer quadruple outages, etc.).

\begin{figure}[tb]
    \centering
    \includegraphics[width=0.7\linewidth]{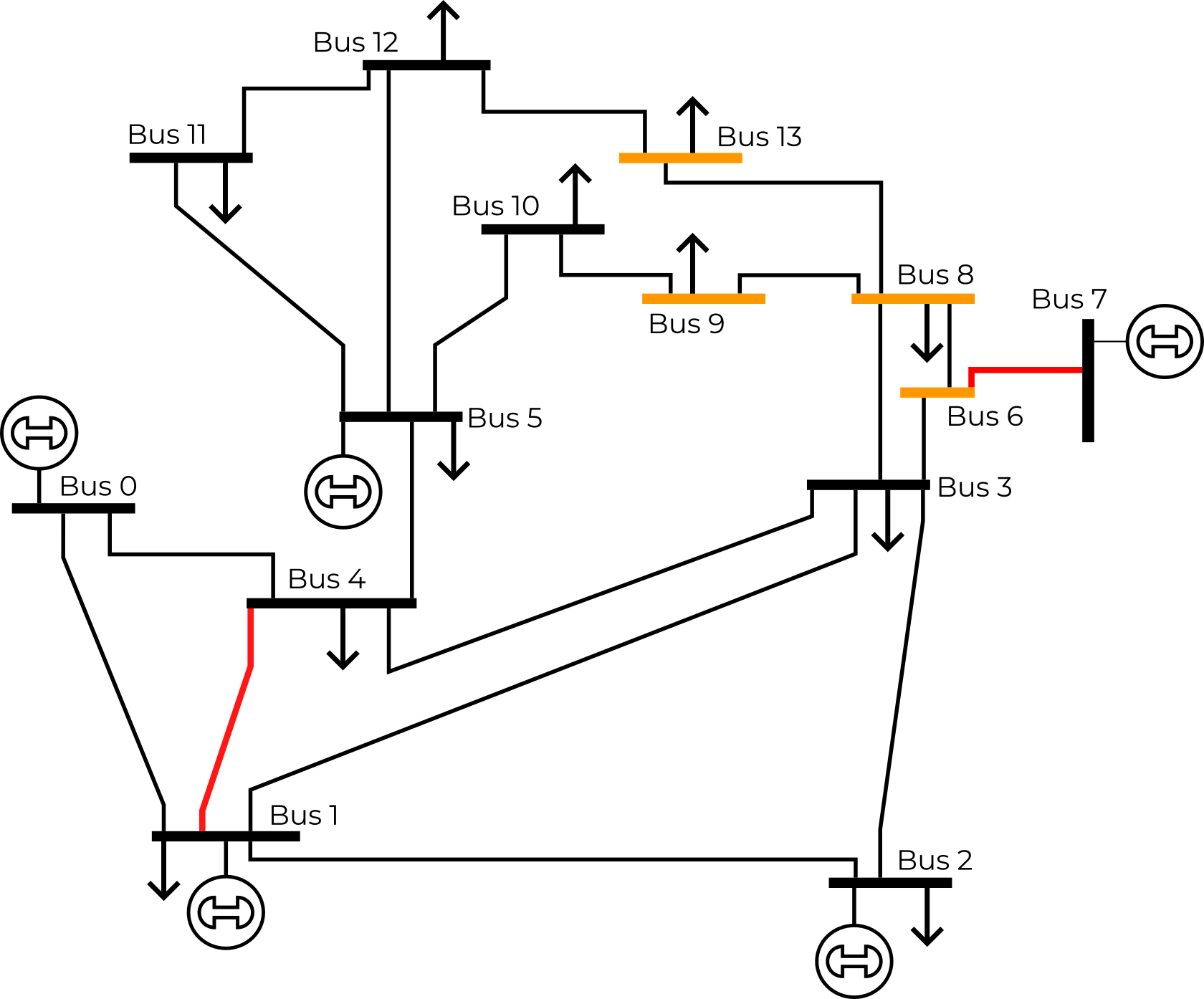}
    \caption{The worst-case outage for the 14-bus test network, predicted using our method. None of the $10^6$ randomly-sampled outages led to worse performance than this double outage, which involved a full loss of the line connecting the generator at bus 7 along with a partial loss in the lower-left.}
    \label{fig:outage}
\end{figure}

Moreover, while 17.7\% of randomly sampled outages were worse (in terms of voltage and power limit violation) than the worst-case contingency predicted by adversarial optimization, none of the randomly-sampled outages were worse than that predicted by our method, which for the 14-bus network was a double failure with complete loss of the line connecting the generator at bus 7 and a partial loss of the line between buses 1 and 4 (as shown in Fig.~\ref{fig:outage}).

\section{Conclusion}

In this paper, we have shown that a common solution method for SCOPF (adversarial optimization) is liable to severely overestimate the robustness of its solution: the contingencies predicted during the adversarial optimization process are not representative of the full space of possible outages. To solve this issue, we re-frame SCOPF as an adversarial sampling problem (rather than optimization) and present a novel algorithm for not only predicting a diverse set of possible contingencies but also updating the dispatch to mitigate those contingencies. Through experiments on small-scale transmission networks, we demonstrate that our approach avoids the false optimism of adversarial optimization (given the same computational budget). Future work involves scaling our approach to practically-sized transmission networks, migrating from our existing Python implementation to C++.

Given the complexity of coordinating generation and load in electricity transmission networks, system operators must rely on semi-automated optimization tools; however, it is important that operators be able to trust the tools they rely on. Our work aims to highlight a potential issue with relying solely on adversarial optimization-based tools for SCOPF (namely, the risk of false optimism) and provide a solution for mitigating these issues and finding more robust generation schedules. We hope that our adversarial sampling framework will inspire a new family of algorithms for this problem, enabling more reliable and secure operation of the electricity grid.

\bibliographystyle{IEEEtran}
\bibliography{IEEEabrv,main}

\end{document}